\RequirePackage[mathlines]{lineno} 
\documentclass[aps,prd,twocolumn,showpacs,amsmath,amssymb,nofootinbib]{revtex4-1}
\usepackage{epsfig}
\usepackage{graphicx}
\usepackage{dcolumn}
\usepackage{bm}
\usepackage{overpic}
\usepackage{subfigure}
\usepackage{float}
\usepackage{color}
\usepackage{amsmath} 
\usepackage{mathcomp}
\usepackage{multirow}
\usepackage{rotating}
\newcommand{\PreserveBackslash}[1]{\let\temp=\\#1\let\\=\temp}
\newcolumntype{C}[1]{>{\PreserveBackslash\centering}p{#1}}
\newcolumntype{R}[1]{>{\PreserveBackslash\raggedleft}p{#1}}
\newcolumntype{L}[1]{>{\PreserveBackslash\raggedright}p{#1}}

\newcommand{\jpsi}{J/\psi}

\begin{document}
\normalsize
\parskip=5pt plus 1pt minus 1pt

\title{ Measurement of {\boldmath $J/\psi\to\Xi(1530)^{-}\bar\Xi^{+}$} and evidence for the
radiative decay {\boldmath $\Xi(1530)^{-}\to\gamma\Xi^-$} }

%

\author{
\begin{small}
\begin{center}
M.~Ablikim$^{1}$, M.~N.~Achasov$^{10,d}$, P.~Adlarson$^{59}$, S. ~Ahmed$^{15}$, M.~Albrecht$^{4}$, M.~Alekseev$^{58A,58C}$, A.~Amoroso$^{58A,58C}$, F.~F.~An$^{1}$, Q.~An$^{55,43}$, Y.~Bai$^{42}$
% ...
\\
\vspace{0.2cm}
(BESIII Collaboration)\\
\vspace{0.2cm} {\it
$^{1}$ Institute of High Energy Physics, Beijing 100049, People's Republic of China\\
% ...
$^{4}$ Bochum Ruhr-University, D-44780 Bochum, Germany\\
% ...
$^{10}$ G.I. Budker Institute of Nuclear Physics SB RAS (BINP), Novosibirsk 630090, Russia\\
% ...
$^{15}$ Helmholtz Institute Mainz, Johann-Joachim-Becher-Weg 45, D-55099 Mainz, Germany\\
% ...
$^{42}$ Southeast University, Nanjing 211100, People's Republic of China\\
$^{43}$ State Key Laboratory of Particle Detection and Electronics, Beijing 100049, Hefei 230026, People's Republic of China\\
% ...
$^{55}$ University of Science and Technology of China, Hefei 230026, People's Republic of China\\
% ...
$^{58}$ (A)University of Turin, I-10125, Turin, Italy; (B)University of Eastern Piedmont, I-15121, Alessandria, Italy; (C)INFN, I-10125, Turin, Italy\\
$^{59}$ Uppsala University, Box 516, SE-75120 Uppsala, Sweden\\
% ...
$^{d}$ Also at the Novosibirsk State University, Novosibirsk, 630090, Russia\\
% ...
}\end{center}
\vspace{0.4cm}
\end{small}
}

\noaffiliation

\vspace{4cm}
\date{\today}

\begin{abstract}
  The SU(3)-flavor violating decay $J/\psi\to\Xi(1530)^{-}\bar\Xi^{+}+c.c.$ is studied using $(1310.6\pm7.0)\times 10^{6} ~J/\psi$ events collected
  with the BESIII detector at BEPCII, and the branching fraction is measured to be
  ${\cal{B}}(J/\psi\to\Xi(1530)^{-}\bar\Xi^{+}+c.c.)=(3.17\pm0.02_{\rm stat.}\pm0.08_{\rm syst.})\times10^{-4}$. This result is
  consistent with previous measurements with an order of magnitude improved precision.
  The angular parameter for this decay is measured  for the first time and is found to be $\alpha=-0.21\pm0.04_{\rm stat.}\pm0.06_{\rm syst.}$.
  In addition, we report  evidence for the radiative decay $\Xi(1530)^{-}\to\gamma\Xi^- $ with a significance of 3.9$\sigma$, including the systematic uncertainties.
  The 90\%  confidence level upper limit  on the branching fraction is determined to be $\mathcal{B}(\Xi(1530)^{-}\to\gamma\Xi^- )\leq3.7$\%.

\end{abstract}


\maketitle

\section{Introduction}
\label{sec:introduction}
\vspace{-0.4cm}

The $\Xi$ and $\Xi(1530)$ hyperons are regarded as SU(3) octet (orbital angular momentum within quarks $L=0$ and spin-parity $J^P=\frac{1}{2}^+$) and decuplet ($L=0$ and $J^P=\frac{3}{2}^+$) baryons, respectively~\cite{su3,jpsidecay1989,Ramalho:2013uza}. In this context, the process $J/\psi\to\Xi(1530)^{-}\bar\Xi^{+}$~\cite{R00} should be suppressed by the SU(3)-flavor symmetry~\cite{su3,jpsidecay1989,Jpsidecay1976}.
Nevertheless, a sizable branching fraction of $(5.9\pm1.5)\times10^{-4}$~\cite{R11,DM2} for the decay $J/\psi\to\Xi(1530)^{-}\bar\Xi^{+}$ was measured based on
$(8.6\pm1.3)\times10^{6}$ $J/\psi$ events by the DM2 Collaboration in 1982, and $(0.70\pm0.12)\times10^{-5}$ for the decay $\psi(3686)\to
\Xi(1530)^{-}\bar\Xi^{+}$ based on $(448.1\pm2.9)\times10^{6}$ $\psi(3686)$ events by the BESIII Collaboration in 2019~\cite{wxf}.
For comparison, the SU(3)-flavor violating decay $J/\psi\to\Delta^{+}\bar{p}$ has a branching fraction of less than $1\times10^{-4}$~\cite{DM2} at $90\%$ confidence level (C.L.), while the SU(3)-allowed decays $J/\psi\to p \bar{p}$ and $J/\psi\to~N(1535)^{+}\bar{p}$~\cite{su3} have branching fractions of $(1-2)\times10^{-3}$~\cite{pdg}.
Therefore, the branching fraction for $J/\psi\to\Xi(1530)^{-}\bar\Xi^{+}$~\cite{DM2} is anomalously large when compared 
to that of $J/\psi\to\Xi^-\bar\Xi^+$, which is measured to be $(0.98\pm0.08)\times10^{-3}$~\cite{pdg}.
An explanation for this anomaly is that a substantial $J^P=\frac{1}{2}^-$ component may hide underneath the $J^P=\frac{3}{2}^+$
 peak while the branching fraction for $J/\psi\to\Xi(1530)^{-}\bar\Xi^{+}$ was obtained assuming a pure $\frac{3}{2}^+$
  contribution around 1530 MeV/$c^2$~\cite{su3}.
An isodoublet $\Xi^*$ state with $J^P= \frac{1}{2}^-$
around 1520 MeV/$c^2$~\cite{Zhang:2004xt}, called $\Xi(1520)$,
is predicted in the diquark cluster picture, which is an SU(3) pentaquark octet with a $[ds][su]\bar{u}$ component.
Due to the small number of event in the analysis of $J/\psi\to\Xi(1530)^{-}\bar\Xi^{+}$
reported by DM2~\cite{DM2}, it is difficult to give a solid conclusion on whether a $\frac{1}{2}^-$ partial wave contributes to the $\Xi(1530)$ mass region.

BESIII collected $(1310.6\pm7.0)\times 10^{6}$~$J/\psi$ events~\cite{data0,totJpsi} in 2009 and 2012, a two orders of magnitude larger statistics than available to the DM2 experiment. A precision measurement with the BESIII experiment was therefore performed.

In 1981, Brodsky and Lepage~\cite{Brodsky:1981kj} were the first to note the significance of angular distributions as a test of quantum chromodynamics.
According to Ref.~\cite{Brodsky:1981kj}, the angular distribution of the $J/\psi$ decay to a baryon-antibaryon ($B\bar{B}$) pair is defined by:
\begin{equation}\label{equ:alpha}
  \frac{dN}{d {\rm cos}\theta}\propto 1 + \alpha {\rm cos}^2\theta,
\end{equation}
where $\theta$ is the polar angle between the baryon direction and the positron beam direction in the $J/\psi$ rest frame,
and $\alpha$ is a constant that parameterizes the angular distribution. The value of $\alpha$ has been predicted in many  theoretical approaches for the SU(3)-allowed charmonium decays, such as electromagnetic contributions~\cite{alpha1}, quark mass effects~\cite{alpha2, alpha3}, rescattering effects~\cite{alpha4}, {\it etc}.
Considering electromagnetic contributions while ignoring quark mass effects in the SU(3)-allowed  $J/\psi \rightarrow B\bar{B}$ decays, the parameter $\alpha$ is expressed~\cite{alpha1} as $$\alpha=\frac{m_{J/\psi}^2-4M_B^2}{m_{J/\psi}^2+4M_B^2},$$
where $m_{J/\psi}$ is the nominal $J/\psi$ mass~\cite{pdg} and $M_B$ refers to a baryon mass. Yet Carimalo~\cite{alpha2}
deemed that quark mass effects are more sensitive than electromagnetic contributions to the $\alpha$ value.
He provied the formula~\cite{alpha2}  $$\alpha=\frac{(1+u)^2-u(1+6u)^2}{(1+u)^2+u(1+6u)^2}$$
with $u=M_B^2/m_\psi^2$ ($m_\psi$ denotes a charmonium resonance mass),
which fits the experimental data better than when
only considering electromagnetic effects. It is easy to see that  $0<\alpha<1$  in the above-mentioned parameterizations.
However, BESIII previously measured a negative $\alpha$ values for $J/\psi \rightarrow \Sigma^0 \bar\Sigma^0$ and $\Sigma(1385)\bar\Sigma(1385)$~\cite{Ablikim:2005cda, xiongfei1617}.
Chen and Ping~\cite{alpha4} investigated the rescattering effects of $B\bar{B}$ in heavy quarkonium decays. As a result, the resulting angular distribution parameter $\alpha$ can be negative.
 However, there are no theoretical predictions or experimental data available on the angular distributions for SU(3)-flavor violating $J/\psi$ decays. Measurements of angular distributions of such decays have the potential to bring more insight into the SU(3)-flavor violating mechanism.

In addition, the electromagnetic transition of decuplet to octet hyperons is a very sensitive probe of their structures \cite{Ramalho:2013uza, c1, Myhrer:2006cu, Li:2016tlt}. The partial width of the radiative transition
$\Xi(1530)^-\to \gamma \Xi^-$  is estimated to be  3.1 keV when considering meson cloud effects with a relativistic quark model~\cite{Ramalho:2013uza} in which the valence quark contributions for a baryon are supplemented by the pion or kaon cloud, and about 3 keV when considering octet-decuplet mixing with a nonrelativistic potential model~\cite{c1}.
Taking into account the total decay width of $\Xi(1530)^-$ of 9.9 MeV~\cite{pdg}, the branching fraction of $\Xi(1530)^{-}\to\gamma\Xi^{-}$ is inferred to be about 3.0$\times 10^{-4}$. Experimentally, only an upper limit for ${\cal{B}}(\Xi(1530)^-\to\gamma\Xi^-)<4\%$
is reported at the 90\% C.L. in 1975 \cite{c2}.

In this analysis, based on $(1310.6\pm7.0)\times 10^{6}$~$J/\psi$ events~\cite{totJpsi} collected with the BEijing Spectrometer III (BESIII) at the Beijing Electron-Positron Collider (BEPCII), we measure the branching fraction of $\jpsi\to\Xi(1530)^{-}\bar{\Xi}^{+}$ with an improved precision and determine the angular distribution parameter for the first time.
In addition, we also report evidence for the $\Xi(1530)^{-}\to\gamma\Xi^-$ decay with a 3.9$\sigma$ significance based on the $J/\psi\to\Xi(1530)^{-}\bar\Xi^{+}$ process,
and the corresponding 90\% C.L. upper limit on the branching fraction is given.

\section{BESIII detector and monte carlo simulation}
\label{sec:BESIII}
\vspace{-0.4cm} The BESIII detector  operating at the BEPCII collider
is described in detail in
Ref.~\cite{nima614.345}.  The detector is cylindrically symmetric and
covers 93\% of $4\pi$ solid angle.  It consists of the following four
sub-detectors: a 43-layer main drift chamber (MDC), which is used to determine
momenta of charged tracks with a resolution of 0.5\% at 1 GeV/$c$
in an axial magnetic field of 1 T with the 2009 dataset and 0.9 T with the 2012 dataset; a plastic scintillator
time-of-flight system (TOF), with a time resolution of 80~ps (110~ps)
in the barrel (endcaps); an electromagnetic calorimeter (EMC)
consisting of 6240 CsI(Tl) crystals, with relative photon energy resolution
of 2.5\% (5\%) at 1 GeV in the barrel (endcaps); and a muon counter
consisting of 9~(8) layers of resistive plate chambers in the barrel
(endcaps), with a position resolution of 2 cm.

The response of the BESIII detector is modeled with Monte Carlo (MC) simulations using the software framework \textsc{boost}~\cite{cpc30.371} based on \textsc{geant}4~\cite{numa506.250,tns53.270}, which includes the geometry and
material description of the BESIII detectors, the detector response
and digitization models, as well as a database that keeps track 
of the running conditions and the detector performance.
MC samples are used to optimize the selection criteria, evaluate the signal efficiency, and
estimate backgrounds.
Two signal MC samples of 0.3 million events each have been generated with the {\sc J2BB3} model~\cite{J2BB3} for the $J/\psi\to\Xi(1530)^-\bar\Xi^+$ reaction.
The first MC sample contains inclusive $\Xi(1530)^-$ decays and the second sample consists of exclusive $\Xi(1530)^-\to\gamma\Xi^-$ decay using the angular distribution constant $\alpha$ (see Eq.~(\ref{equ:alpha}) of Ref.~\cite{J2BB3}) as measured in this analysis.
Only the baryon decays $\bar\Xi^+~\to \bar\Lambda \pi^+$ and $\bar\Lambda~\to \bar{p} \pi^+$ in the signal
channels are simulated.
An inclusive MC sample of $1.225\times10^9$ $J/\psi$ events is used for the background studies.
Here, the $J/\psi$ resonance is produced by means of the \textsc{kkmc} event generator \cite{kkmc}, in which the initial state radiation is
included. The decays are simulated by {\sc evtgen}~\cite{evtgen} with
the known branching fractions taken from the Particle Data Group (PDG)~\cite{pdg},
while the remaining unmeasured decay modes are generated with \textsc{lundcharm}~\cite{lundcharm}.

\section{Data Analysis}
\label{sec:selection}
\subsection{ {\boldmath $J/\psi\to\Xi(1530)^-\bar\Xi^+$} with  {\boldmath $\Xi(1530)^-\to$} anything}
\label{sec:select1}

For the inclusive analysis of the $\Xi(1530)^-$ decay,
a single tagged (ST) $\bar\Xi^+$ baryon candidate
is reconstructed via $\bar\Lambda(\to \bar{p} \pi^+)\pi^+$,
while the $\Xi(1530)^-$ candidate is treated as a missing particle.
The presence of a $\Xi(1530)^-$ candidate is inferred using the mass
recoiling against the $\bar\Lambda\pi^+$ system, $M^{\rm recoil}_{\bar\Lambda\pi^+}=\sqrt{(E-E_{\bar\Lambda\pi^+})^2-(\bm{P}_{\bar\Lambda\pi^+})^2 }$, where $E$ is the center-of-mass (c.m.) energy and $(E_{\bar\Lambda\pi^+}, \bm{P}_{\bar\Lambda\pi^+})$
is the four momenta of the $\bar\Lambda\pi^+$ system in the $e^+e^-$ rest frame.
For signal candidate events, the distribution of $M^{\rm recoil}_{\bar\Lambda\pi^+}$ will form a peak around the
nominal mass of the charged $\Xi(1530)^-$ resonance~\cite{pdg}.

Charged tracks must be properly reconstructed in the MDC with $|{\rm cos}\theta|~<~0.93$, where
$\theta$ is the polar angle between the charged track and the positron beam direction.
The combined information from the TOF and ionization loss ($dE/dx$) in the MDC is used to calculate particle identification confidence levels for each hadron ($i$) hypothesis ($i = p, \pi, K$).
A charged track is identified as the $i-$th particle type with the highest confidence level.
Events with at least one antiproton (proton) and two positively (negatively) charged pions are selected for tagging the $\bar\Xi^+$ ($\Xi^-$) decay mode.

\begin{figure*}[htbp]
\centering
  \mbox{
    \begin{overpic}[width=0.32\textwidth,clip=true]{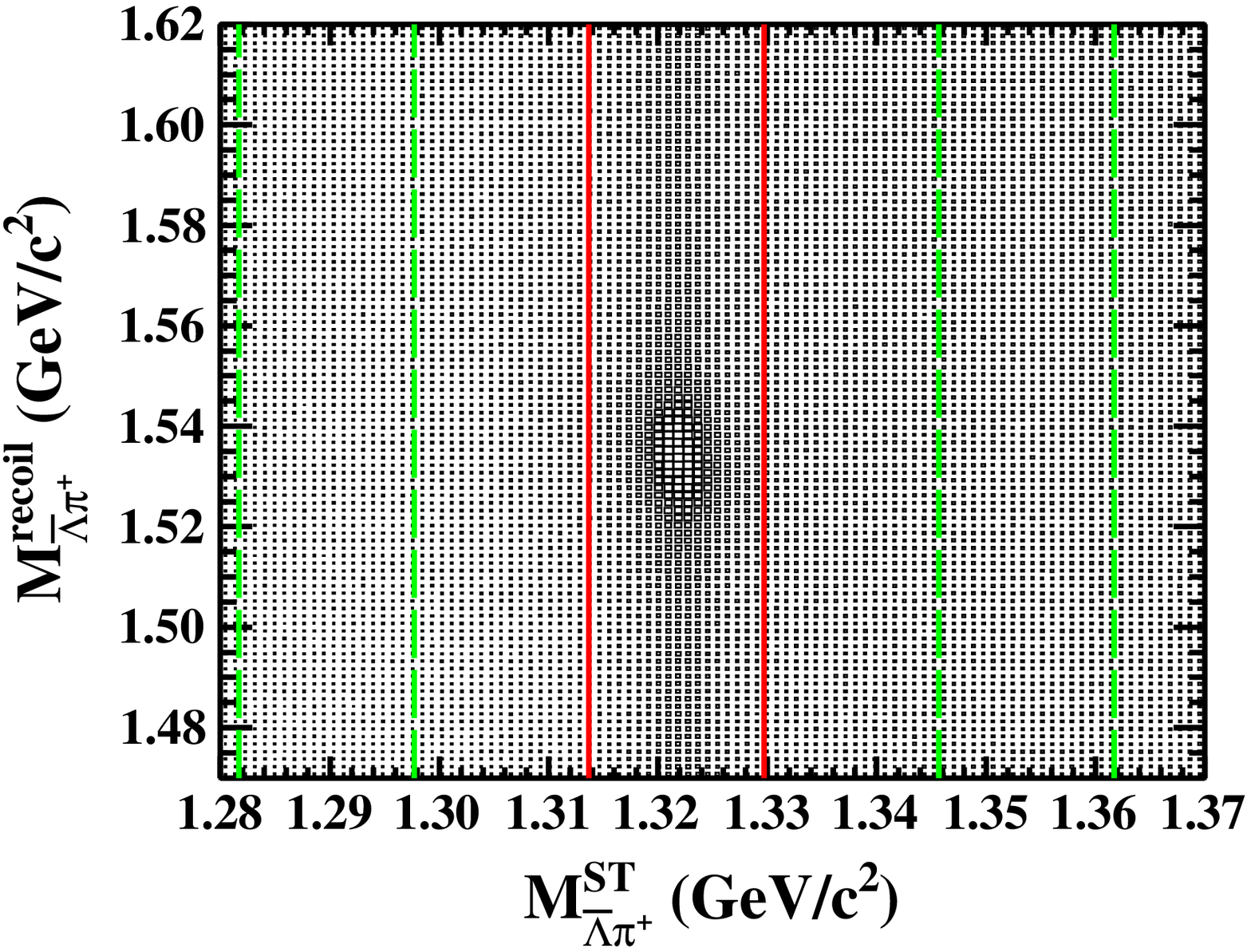}
    \end{overpic}
    \begin{overpic}[width=0.32\textwidth,clip=true]{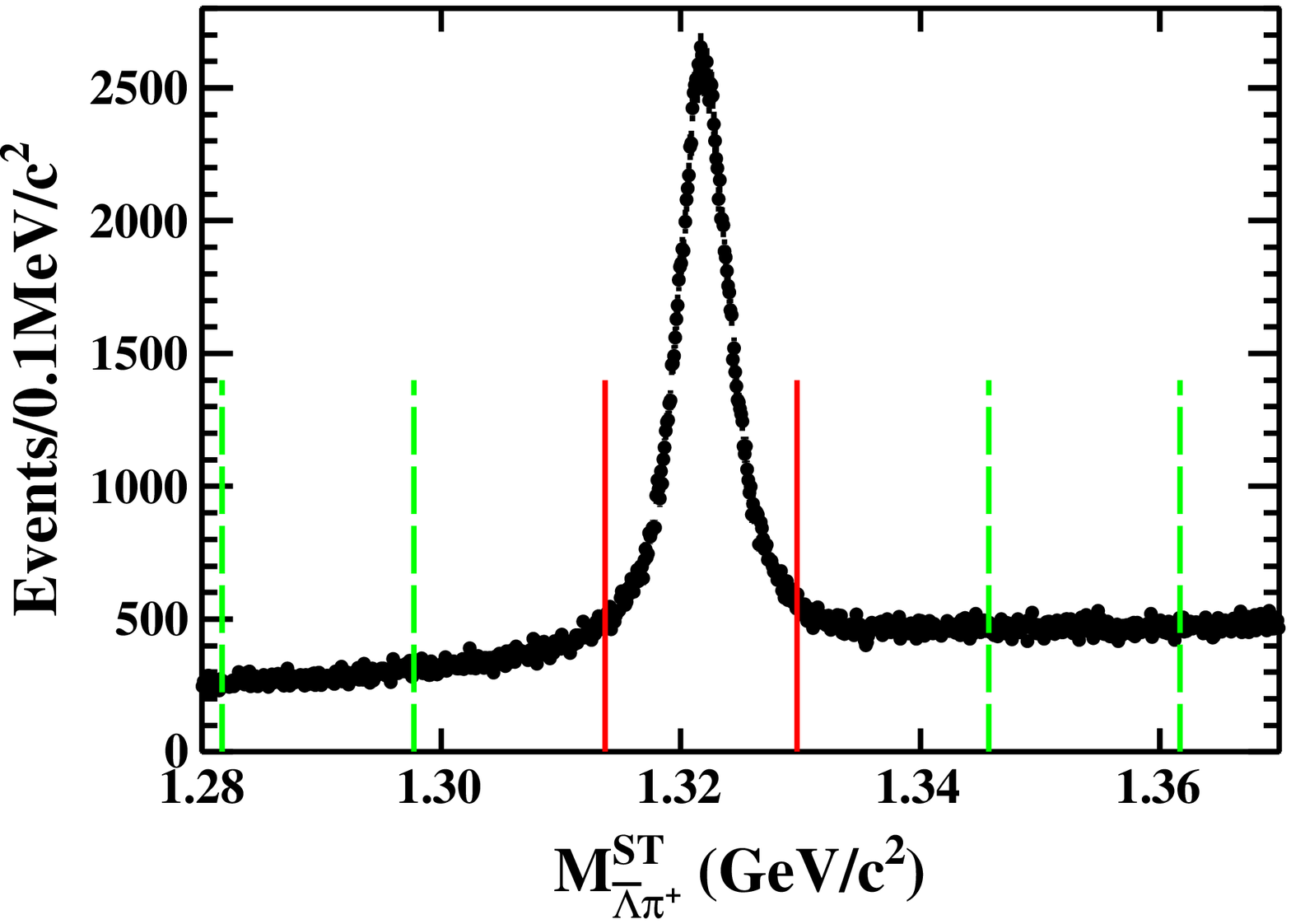}
    \end{overpic}
    \begin{overpic}[width=0.32\textwidth,height=4.4cm,clip=true]{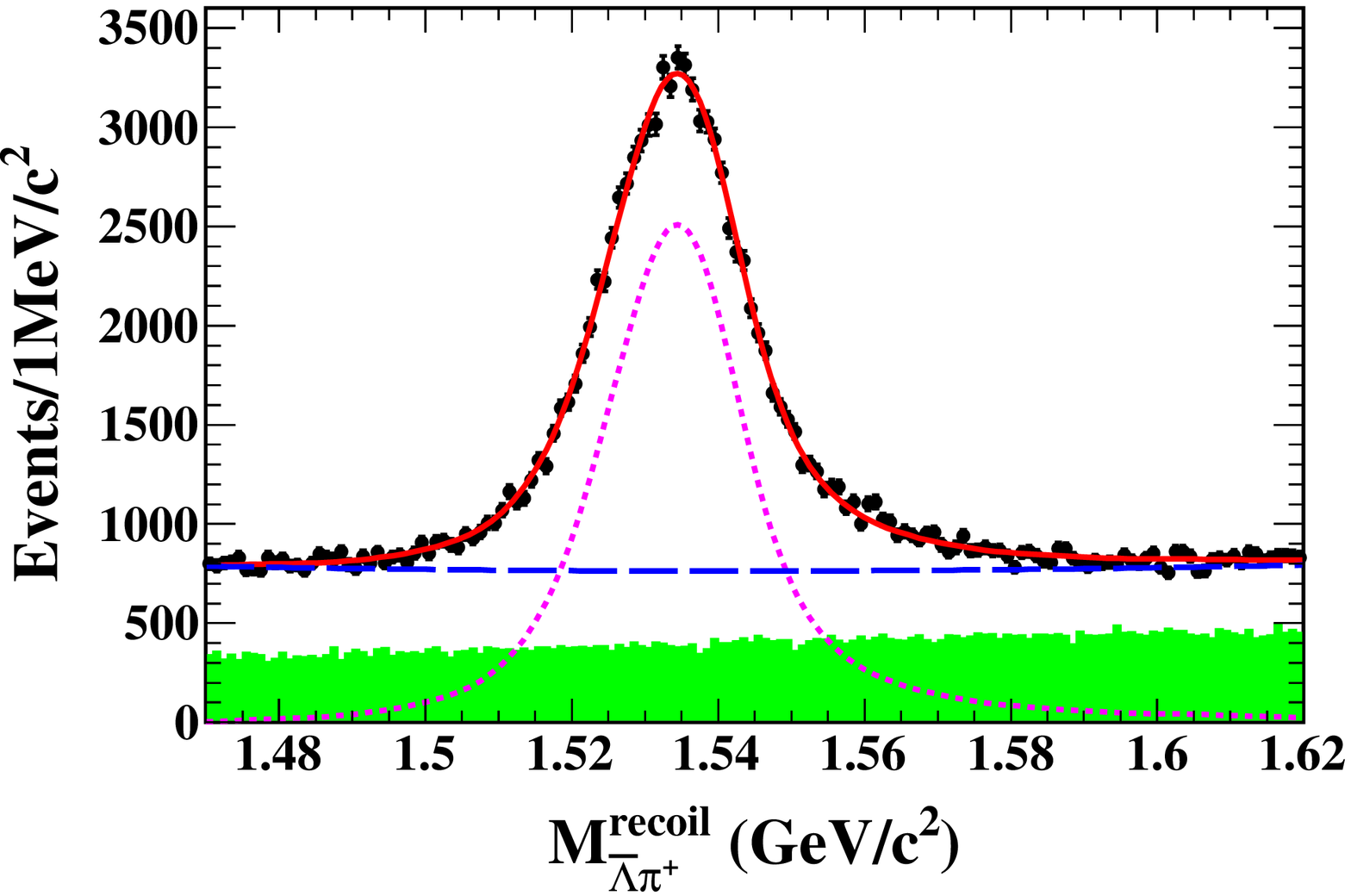}
    \end{overpic}
   }
\caption{Left: scatter plot of $M^{\rm recoil}_{\bar\Lambda\pi^{+}}$ versus $M^{\rm ST}_{\bar{\Lambda}\pi^{+}}$
  from the data, where $M^{\rm ST}_{\bar{\Lambda}\pi^{+}}$ is the $\bar{\Lambda}\pi^{+}$ invariant mass in the ST mode.
  Middle: the $M^{\rm ST}_{\bar{\Lambda}\pi^{+}}$ distribution in the data.
  The red solid and green long-dashed lines indicate the $\bar\Xi^+$ signal and sideband regions, respectively.
  Right: fit to the experimental $M^{\rm recoil}_{\bar{\Lambda}\pi^{+}}$ distribution.
  The red solid line is the fit result, the pink dotted line denotes the signal component, the blue long-dashed line represents the fitted background component,
  and the green-shaded histogram represents the normalized $\bar\Xi^+$ mass sideband events from the data.
  }
\label{scatter}
\end{figure*}

The $\bar\Lambda$ candidates are reconstructed with a vertex fit to all the identified $\bar{p}\pi^+$ combinations.
A secondary vertex fit~\cite{sec-vtx} is then employed to the $\bar\Lambda$ candidates and events are kept if the decay length, $i.e.$ the distance from the production vertex to the decay vertex, is greater than zero.
If there remains more than one $\bar{p}\pi^+$ combination in the event, the one closest
to the nominal $\bar\Lambda$ mass~\cite{pdg} is retained.
A $\bar\Lambda$ signal is required to have a $\bar{p}\pi^+$ invariant mass
within 5 MeV$/c^2$ from the nominal $\bar\Lambda$ mass~\cite{pdg}.
The $\bar\Xi^+$ candidates are reconstructed via a secondary vertex fit
by considering all combinations of the extra charged pions and the selected $\bar\Lambda$ candidate,
requiring that the decay length of the reconstructed $\bar\Xi^+$ candidates are greater than zero.
If several combinations remain, the one with the minimum $|M_{\bar\Lambda\pi^+}-m_{\bar\Xi^+}|$, where $M_{\bar\Lambda\pi^+}$ is the invariant mass of the $\bar\Lambda\pi^+$ system and $m_{\bar\Xi^+}$ is the nominal mass of the $\bar\Xi^+$ baryon~\cite{pdg}, is selected.
Additionally, the requirement $|M_{\bar\Lambda\pi^+}-m_{\bar\Xi^+}|\leq8 $ MeV/c$^{2}$ is
applied to further suppress the backgrounds.

After applying the above selection criteria,
a scatter plot of $M^{\rm recoil}_{\bar\Lambda\pi^{+}}$ versus $M^{\rm ST}_{\bar{\Lambda}\pi^{+}}$ is shown in Fig.~\ref{scatter}(left), where $M^{\rm ST}_{\bar{\Lambda}\pi^{+}}$ is the $\bar{\Lambda}\pi^{+}$ invariant mass in the ST mode,
and significantly clustered events of the SU(3)-flavor violating $J/\psi \to \Xi(1530)^-\bar{\Xi}^+$ decay are observed in the data.
Figure~\ref{scatter}(middle) illustrates the distribution of $M^{\rm ST}_{\bar{\Lambda}\pi^{+}}$.
In both figures, the red solid and green long-dashed lines indicate the $\bar\Xi^+$ signal and sideband regions, respectively.
The $\Xi(1530)^-$ signal in the $M^{\rm recoil}_{\bar\Lambda\pi^+}$ spectrum has a Breit-Winger shape, as shown in Fig.~\ref{scatter}(right).

The continuum data collected at the c.m. energy of 3.08~GeV, with an integrated luminosity of 30~pb$^{-1}$~\cite{data0,totJpsi}, are used to investigate the contribution from the quantum electrodynamics (QED) process $e^{+}e^{-}\to\Xi(1530)^-\bar{\Xi}^+$. By imposing the same event selection criteria as the $J/\psi$ data{\color{red},} 
no events survived, meaning that the QED background is negligible.
The contamination from the non-$\bar\Xi^+$ backgrounds is estimated with the $\bar\Xi^+$ mass sideband events, where the sideband regions are selected as $M^{\rm ST}_{\bar{\Lambda}\pi^{+}}\in[1.2817, 1.2977]\cup[1.3457, 1.3617]$ GeV$/c^2$, as
indicated by the green long-dashed lines in Fig.~\ref{scatter}(middle). No peaking background is found
in the $\Xi(1530)^-$ signal region from the $\bar\Xi^+$ mass sideband events, as
indicated by the green-shaded histogram in Fig.~\ref{scatter}(right). The remaining backgrounds, investigated by the inclusive MC sample, form a smooth distribution in the $M^{\rm recoil}_{\bar\Lambda\pi^+}$ spectrum in the region of 1.535 GeV$/c^2$, where the main contributions are
from $J/\psi\to\Xi^{-}\bar{\Xi}^{+}\pi^0$ and $J/\psi\to\Xi^{0}\bar{\Xi}^{+}\pi^{-} $ events.

\begin{table}[!hbp]
\caption{Numerical results on the branching fraction measurement for $J/\psi\to\Xi(1530)^-\bar\Xi^+$. The uncertainties are statistical only.
}
\begin{center}
\renewcommand\arraystretch{1.2} 
\begin{tabular*}{\columnwidth}{@{\extracolsep{\fill}}cc}
\hline
\hline
$N^{\rm obs}_{\rm ST}$ & $70186\pm544$ \\ 
$N_{J/\psi}$ & $1310.6\times 10^6$ \\  
${\cal{B}}(\bar\Xi\to\bar\Lambda\pi^+)$   &  99.89\% \\
${\cal{B}}(\bar\Lambda\to \bar{p}\pi^+)$     & 63.90\%    \\
$\epsilon^{\Xi^{-}}_{\rm ST}$ ($\epsilon^{\bar{\Xi}^{+}}_{\rm ST}$)  &24.03\% (25.57\%)    \\
$f^-$ ($f^+$)  & $1.079\pm0.011$ ($1.053\pm0.011$)\\
Branching fraction~($\times10^{-4}$)    &  $3.17\pm0.02$   \\
\hline
\hline
\end{tabular*}
\end{center}
\label{tab:st}
\end{table}

The signal yields of the $J/\psi\to\Xi(1530)^-\bar\Xi^+$ decay are extracted
from an unbinned maximum likelihood fit to the $M^{\rm recoil}_{\bar\Lambda\pi^+}$ spectrum.
The $\Xi(1530)^-$ signal is described by the simulated MC shape convolved with a Gaussian function,
which accounts for the mass resolution difference between the data and MC simulation.
The mean of the Gaussian function is fixed to zero while the standard deviation is a free parameter.
The background contribution is described by a second-order Chebychev polynomial function.
The fit of the $M^{\rm recoil}_{\bar\Lambda\pi^+}$ spectrum in data is shown in Fig.~\ref{scatter} (right),
and the fitted signal yields are listed in Table~\ref{tab:st}.

\subsection{ {\boldmath $J/\psi\to\Xi(1530)^-\bar\Xi^+$} with \boldmath{$\Xi(1530)^-\to\gamma\Xi^-$}}

\begin{figure*}[hbtp]
\centering
  \mbox{
    \begin{overpic}[width=0.32\textwidth,clip=true]{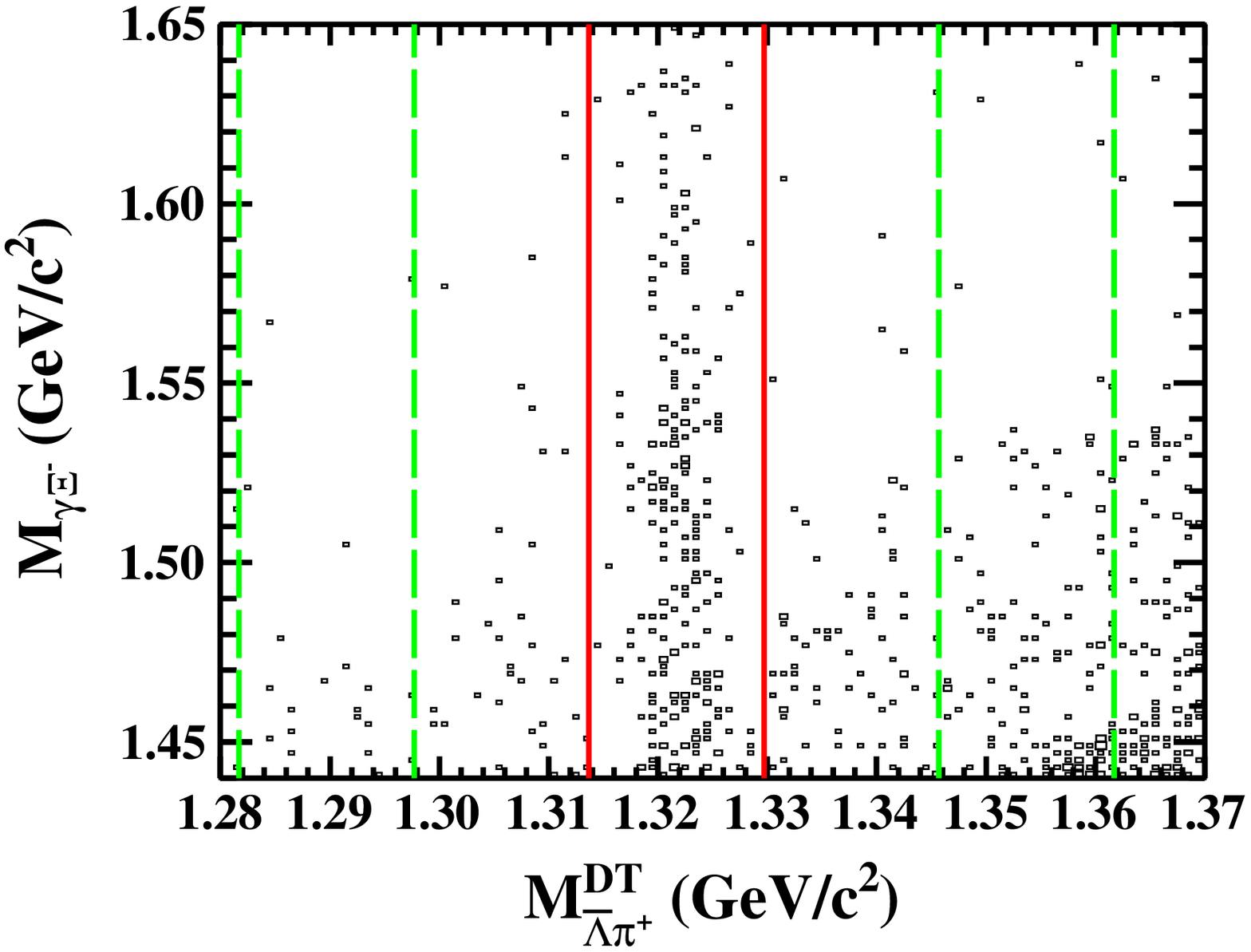}
    \end{overpic}
    \begin{overpic}[width=0.32\textwidth,clip=true]{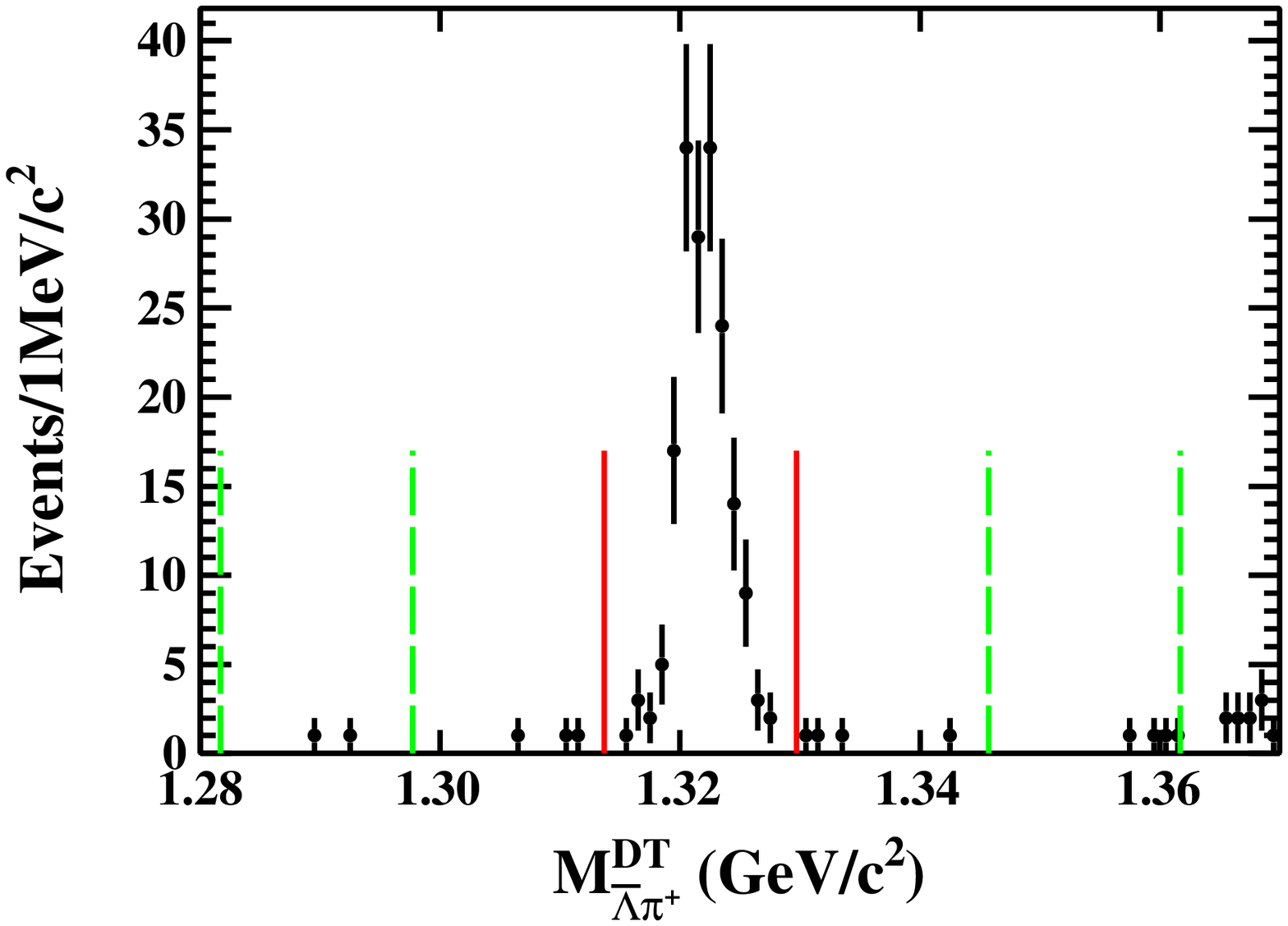}
    \end{overpic}
    \begin{overpic}[width=0.32\textwidth, height=4.3cm, clip=true]{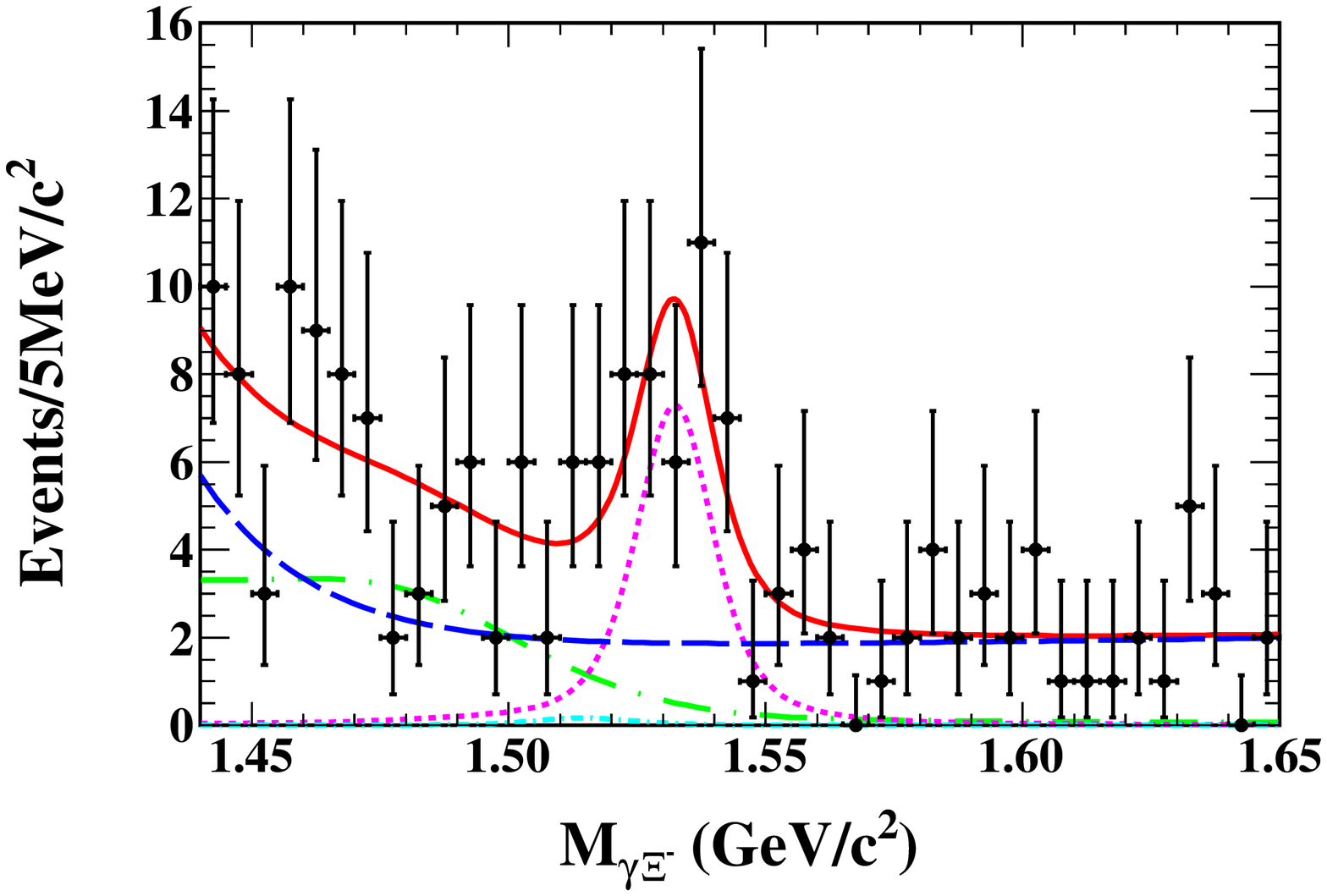}
    \end{overpic}
   }
\caption{Left: scatter plot of $M_{\gamma\Xi^{-}}$ versus $M^{\rm DT}_{\bar{\Lambda}\pi^{+}}$
  from the data, where $M^{\rm DT}_{\bar{\Lambda}\pi^{+}}$ is the $\bar\Lambda\pi^{+}$ invariant mass spectrum in the DT mode.
  Middle: the $M^{\rm DT}_{\bar{\Lambda}\pi^{+}}$ distribution from the data.
  The red solid and green long-dashed lines indicate the $\bar\Xi^+$ signal and sideband regions, respectively.
  Right: the fit to the experimental $M_{\gamma\Xi^{-}}$ distribution.
  The ed solid line is the fit result, the pink dotted line denotes the signal component,
  the cyan dash-dotted line describes the few peaking background events
  from the process $J/\psi\to\Xi(1530)^-\bar\Xi^+$ with $\Xi(1530)^-$ decaying to the $\Xi^-\pi^0$ and $\Xi^0\pi^-$ systems,
  the green long-dash-dotted line denotes the background events from $J/\psi\to\gamma\eta_{c}\to\gamma\Xi^{-}\bar{\Xi}^{+}$,
  and the blue long-dashed line denotes the contribution from the remaining  background events.
  }
\label{scatter2}
\end{figure*}

The event selection criteria for the radiative decay $\Xi(1530)^-\to\gamma\Xi^- $ are based on the $\bar\Xi^{+}$ tagging mode.
Besides the tagged $\bar\Xi^+$ candidates described in Sec.~\ref{sec:select1}, an extra $\Xi^-$ baryon and a photon
are selected to reconstruct the $\Xi(1530)^-$ candidate.
Since all decay particles from $\Xi(1530)^-$ and $\bar\Xi^+$ are reconstructed from the $J/\psi\to\Xi(1530)^-\bar\Xi^+$ process,
it is referred to as the double tag (DT) mode.
The event selection of $\Xi^-$ candidates is similar to those of
tagged $\bar\Xi^+$ candidates in Sec.~\ref{sec:select1},
except for the charge-conjugated final states.
The $\Xi^-$ candidate with the minimum $|M^{\rm DT}_{\Lambda\pi^-} - m_{\Xi^-}|$ is the only one retained,
and then is requirement $|M^{\rm DT}_{\Lambda\pi^-}-m_{\Xi^-}|\leq8 $ MeV/c$^{2}$ applied.
The $\Xi^{-}$ mass window is shown by the red solid lines in Fig.~\ref{scatter2} (left and middle),
where $M^{\rm DT}_{\Lambda\pi^-}$ is the invariant mass of the $\Lambda\pi^-$ system in the DT mode, and $m_{\Xi^-}$ is the nominal mass of the $\Xi^-$ baryon~\cite{pdg}.

Photons are reconstructed by clustering the EMC crystals' signals, and the energy deposited in the nearby TOF counter is included to improve the reconstruction efficiency and energy resolution~\cite{nima614.345}.
A photon candidate is defined as a shower with an energy deposit of at least 25 MeV in the barrel region ($|{\rm cos}\theta<0.8|$) or of at least 50 MeV in the end-cap region ($|0.86 < {\rm cos}\theta| < 0.92$).
Showers in the angular range between the barrel and the endcaps are poorly reconstructed and therefore excluded. An additional requirement on the EMC timing of a photon candidate, $0\leq t \leq700$~ns, is employed
to suppress electronic noise and energy deposits unrelated to the collision event, where time is measured relative to the event start time. All photons, which satisfy the above selection criteria are kept for further analysis.

A four-constraint (4C) kinematic fit is performed for events with $\gamma$, $\Xi^-$, and $\bar\Xi^+$ candidates by imposing overall energy-momentum conservation.
For each event, the combination with the lowest $\chi^2_{\rm4C}$ is selected.
To suppress background events different from the final states of the signal channel,
we require $\chi^2_{\rm4C}<5$,
which is determined by maximizing the figure-of-merit FOM=$S/\sqrt{S+B}$. Here,
$S$ is the expected number of signal events from the signal MC simulation and $B$ is the number of background events from the inclusive MC sample in which the main background processes (see below in the section) are known and normalized using PDG branching fraction values~\cite{pdg}. Three iterations between the $S$ value and the $\chi^2_{\rm 4C}$ requirement are employed until the procedure is converged.


The $\gamma\Xi^-$ invariant mass spectrum of the events that remain after imposing the selection criteria above are shown in Fig.~\ref{scatter2}~(right). A weak enhancement of events in the region of the radiative $\Xi(1530)^-$ decay can be seen.

The background sources are divided into two categories, one with and one without the $\Xi^-$ resonance.
~The non-$\Xi^-$  backgrounds are investigated by the $\Xi^-$ mass sideband events,
where the sideband regions are defined as in the ST mode (see Sec.~\ref{sec:select1}).
It is found that very few events from the sidebands survived in the $M_{\gamma\Xi^-}$ region around 1.535 GeV$/c^2$.
According to the inclusive MC information, the main background is the decay $J/\psi\to\gamma\eta_{c}\to\gamma\Xi^{-}\bar{\Xi}^{+}$,
which distributes smoothly in the signal region of the $\Xi(1530)^-$ baryon.
Only a few peaking background events contributing to the $\Xi(1530)$ mass region are found from the process $J/\psi\to\Xi(1530)^-\bar\Xi^+$
with $\Xi(1530)^-$ decaying to the $\Xi^-\pi^0$ and $\Xi^0(\to\Lambda\pi^0)\pi^-$ systems
with a soft photon being undetected.
Other background events, forming a flat distribution in the $\gamma \Xi^-$ mass spectrum, arise from the decays $J/\psi\to\gamma\Xi^{-}\bar{\Xi}^{+}$ and $J/\psi\to\Xi^{-}\bar{\Xi}^{+}$.

The signal yields for the decay $J/\psi\to\Xi(1530)^-\bar\Xi^+\to\gamma\Xi^-\bar{\Xi}^+$
are extracted by an unbinned maximum likelihood fit to the $M_{\gamma\Xi^-}$ spectrum.
The shape of the invariant mass distribution of the $\Xi(1530)^-$ baryon is modeled based on the prediction of the simulation.
The few peaking background events from the process $J/\psi\to\Xi(1530)^-\bar\Xi^+$,
with $\Xi(1530)^-$ decaying to the $\Xi^-\pi^0$ and $\Xi^0(\to\Lambda\pi^0)\pi^-$ systems,
are normalized with their branching fractions,
where $\mathcal{B}(J/\psi\to\Xi(1530)^-\bar\Xi^+)$ is obtained from this work
and the branching fractions of two $\Xi(1530)^-$ decays are from the PDG~\cite{pdg}.
The smooth and dominating background from $J/\psi\to\gamma\eta_{c}\to\gamma\Xi^{-}\bar{\Xi}^{+}$ events
is described by the MC-determined shape,
where the corresponding number~\cite{pdg} of the background events is normalized to the data.
The remaining background shape is parametrized by an exponential function plus a first-order polynomial
to describe the inclined flat slope in the $M_{\gamma\Xi^-}$ distribution
from the two main backgrounds, $J/\psi\to\gamma\Xi^{-}\bar{\Xi}^{+} $ and $J/\psi\to\Xi^{-}\bar{\Xi}^{+}$.
The parameters of the exponential function and the first-order polynomial are fitted.
The fit, shown in Fig.~\ref{scatter2}(right), yields $33.2\pm9.6$ signal events with a significance of 3.9$\sigma$ which is the most conservative one among various fit scenarios (i.e., different fit range, signal shape, background shape, and background size).
The significance is calculated from the test-statistic
$\sqrt{-2\ln(\mathcal{L}_0/\mathcal{L}_{\rm max})}$ assuming Wilk's theorem~\cite{Wilk}, where
$\mathcal{L}_{\rm max}$ and $\mathcal{L}_0$ are the likelihoods of the
fits with and without the $\Xi(1530)^-$ signal included, respectively.
The upper limit on the signal yield is determined by convolving the likelihood distribution with a Gaussian function with a standard deviation of $\sigma=x\times \Delta$, where $x$ is the number of fitted signal events, and $\Delta$ refers to the total  systematic uncertainty (4.9\%, see Table~\ref{error3}). It is found to be $N^{\rm UL}_{\rm DT}$ = 46 at the 90\% C.L.

\section{Measurements of Branching Fractions and Angular Distribution}

\subsection{Measurements of {\boldmath $\mathcal{B}(J/\psi\to\Xi(1530)^-\bar\Xi^+)$}
and {\boldmath $\mathcal{B}(\Xi(1530)^-\to\gamma\Xi^-)$}}
\label{sec:branch}
The branching fraction for $J/\psi\to\Xi(1530)^-\bar\Xi^+$ is calculated using
\begin{equation}
          {\cal{B}}(J/\psi\to\Xi(1530)^-\bar\Xi^+) = \frac{N^{\rm obs}_{\rm ST}}{ N_{J/\psi}
           {\cal{B}}(\bar\Xi^+){\cal{B}}(\bar\Lambda) \epsilon_{\rm ST}},
\end{equation}
where $N^{\rm obs}_{\rm ST}$ is the number of events for ST, which is extracted
from the fit to  $M^{\rm recoil}_{\bar\Lambda\pi^+}$ spectrum; $N_{J/\psi}$ is the total number of $J/\psi$ events~\cite{totJpsi}; ${\cal{B}}(\bar\Xi^+)$ and ${\cal{B}}(\bar\Lambda)$ are the branching fractions~\cite{pdg} of $\bar\Xi^+\to\bar\Lambda\pi^+$ and $\bar\Lambda\to \bar{p}\pi^+$, respectively;
$\epsilon_{\rm ST}$, expressed as $(f^{+}\epsilon^{\bar{\Xi}^{+}}_{\rm ST}+f^{-}\epsilon^{\Xi^{-}}_{\rm ST})/2$, is the average detection efficiency in the ST mode for both the charge-conjugate processes, where $\epsilon^{\bar{\Xi}^{+}}_{\rm ST}~(\epsilon^{\Xi^{-}}_{\rm ST})$ denotes the MC-simulated efficiency for only tagging $\bar\Xi^+ ~(\Xi^-)$ decay mode, and $f^+ ~(f^-)$ is the correction factor for the $\bar\Xi^{+}$ ($\Xi^{-}$) reconstruction efficiency estimated by using a control sample of $J/\psi\to\Xi^{-}\bar{\Xi}^{+}$ with all polarization parameters considered.
 Here, $f^+ ~(f^-)$ is the ratio of the $\bar\Xi^+$  ($\Xi^{-}$) reconstruction efficiency in the data [$\epsilon^{\bar\Xi^+}_{\rm data}$~($\epsilon^{\Xi^-}_{\rm data}$)] to that in the MC sample  [$\epsilon^{\bar\Xi^+}_{\rm MC}$~($\epsilon^{\Xi^-}_{\rm MC}$)], $i.e.$,  $f^+=\epsilon^{\bar\Xi^+}_{\rm data} / \epsilon^{\bar\Xi^+}_{\rm MC}$ ($f^-=\epsilon^{\Xi^-}_{\rm data} / \epsilon^{\Xi^-}_{\rm MC}$).
 As a result, the branching fraction of ${\cal{B}}(J/\psi\to\Xi(1530)^-\bar\Xi^+)$ is determined to be $(3.17 \pm0.02)\times10^{-4}$ where the uncertainty is statistical only, and other numerical values are listed in Table~\ref{tab:st}.

The upper limit at the 90\% C.L. on the branching fraction for the radiative decay $\Xi(1530)^-\to\gamma\Xi^-$
is calculated using
\begin{equation}
   \label{equ:rad}
   \begin{split}
    & \mathcal{B}^{\rm UL}(\Xi(1530)^-\to\gamma\Xi^-)  \\
    &=\frac{ N^{\rm UL}_{\rm DT} }{ N_{J/\psi} \mathcal{B}(J/\psi\to\Xi(1530)^-\bar\Xi^+){\cal{B}}(\bar\Xi^+){\cal{B}}(\bar\Lambda) {\cal{B}}(\Xi^-){\cal{B}}(\Lambda)  \epsilon_{\rm DT}   }\\
    &=\frac{ N^{\rm UL}_{\rm DT} \epsilon_{\rm ST}}{ {\cal{B}}(\Xi^-){\cal{B}}(\Lambda) N^{\rm obs}_{\rm ST} \epsilon_{\rm DT} },
   \end{split}
 \end{equation}
where  $N^{\rm UL}_{\rm DT}$ is the upper limit on the number of fitted $\Xi(1530)^-\to\gamma\Xi^-$ signal events at  the 90\% C.L.;
${\cal{B}}(\Xi^-)$ and ${\cal{B}}(\Lambda)$ are the branching fractions~\cite{pdg} of $\Xi^-\to\Lambda\pi^-$ and $\Lambda\to p\pi^-$, respectively; $\epsilon_{\rm DT}$, expressed as $f^- f^+ \epsilon^{\rm MC}_{\rm DT}$, is the detection efficiency in the DT mode, where $\epsilon^{\rm MC}_{\rm DT}$ denotes the MC-simulated efficiency using the J2BB3 model~\cite{J2BB3}.
Taking the systematic uncertainty (see Sec.~\ref{sec:syserr}A) into consideration,
the upper limit at the 90\% C.L. on the branching fraction of $\Xi(1530)^-\to\gamma\Xi^-$
is calculated to be 3.7\%.

\subsection{Measurement of the angular distribution in {\boldmath $J/\psi\to\Xi(1530)^-\bar\Xi^+$}}
\label{sec:alpha}

\begin{figure}[hbtp]
\centering
\epsfig{width=0.33\textwidth,clip=true,file=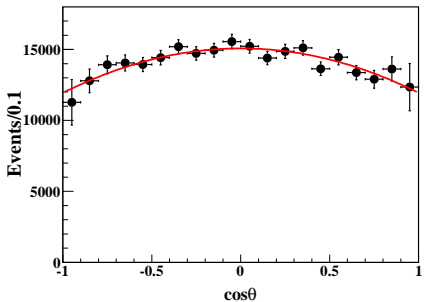}
\caption{The cos$\theta$ distribution for $J/\psi\to\Xi(1530)^-\bar\Xi^+$. The dots with error
bars denote the efficiency-corrected data and the red curve is the fit result.}
\label{angular}
\end{figure}

We obtain the number of recorded $J/\psi \to \Xi(1530)^-\bar\Xi^+$ events in each ${\rm cos}\theta$
bin by fitting the $\bar{\Lambda}\pi^+$ invariant mass distribution as described in Sec.~\ref{sec:branch}.
By dividing by the detection efficiency in each ${\rm cos}\theta$ interval, we obtain the efficiency-corrected  ${\rm cos}\theta$ distribution  shown in Fig.~\ref{angular}.
A least square fit of Eq.~\ref{equ:alpha} to the obtained ${\rm cos}\theta$ distribution in the range of [$-1.0$, 1.0] gives $\alpha=-0.20\pm 0.04$, where the uncertainty is statistical only.

\section{Systematic Uncertainties}
\label{sec:syserr}
\subsection{Branching fractions}
The systematic uncertainties in the branching fraction measurements arise from many sources.
They depend on the $\bar\Xi^+$ efficiency correction,
mass windows for $\bar\Lambda$ and $\bar\Xi^+$, decay lengths for $\bar\Lambda$ and $\bar\Xi^+$,
background shape, the amount of background, the branching fractions of the intermediate decays,
and the total number of $J/\psi$ events. It is noteworthy that the uncertainties due to
the tracking and PID efficiencies for the charged $\pi$ track from the $\bar\Xi^+$ decay
and the $\bar\Lambda$ reconstruction efficiency are included in the charged $\bar\Xi^+$ reconstruction uncertainty.
For the radiative $\Xi(1530)^-$ decay they depend, in addition, on the photon reconstruction efficiency.

\indent 1. Photon reconstruction efficiency:
The uncertainty on the photon detection efficiency is 1.0$\%$ per photon, obtained by studying $J/\psi\to\rho^{0}\pi^{0},
\rho^{0}\to\pi^{+}\pi^{-},\pi^{0}\to\gamma\gamma$ events~\cite{gamrecon}.

\indent 2. $\bar\Xi^+$ efficiency correction:
As mentioned above, the correction factor  $f^+ ~(f^-)$ on the $\bar\Xi^+$ ($\Xi^-$) reconstruction efficiency,
defined as $\epsilon^{\bar\Xi^+}_{\rm data} / \epsilon^{\bar\Xi^+}_{\rm MC}$ ($\epsilon^{\Xi^-}_{\rm data} / \epsilon^{\Xi^-}_{\rm MC}$),
is obtained by using a control
sample of $J/\psi\to \bar\Xi^+ \Xi^-$ decays via single and double tag methods (the values are listed in Table~\ref{tab:st}).
The uncertainty on $f^+ ~(f^-)$, obtained by adding the relative uncertainties for
$\epsilon^{\bar\Xi^+}_{\rm data}$ and $\epsilon^{\bar\Xi^+}_{\rm MC}$ ($\epsilon^{\Xi^-}_{\rm data}$ and $\epsilon^{\Xi^-}_{\rm MC}$) in quadrature assuming the sources are independent,
is found to be 1.0\% for each mode.
Therefore, the systematic uncertainty for  $\bar\Xi^+$ efficiency correction is taken as 0.7\% by averaging both charge-conjugate modes.

\indent 3. Mass window (decay length) of $\bar\Lambda$ ($\bar\Xi^+$):
The uncertainty attributed to the $\bar\Lambda$ ($\bar\Xi^+$) mass window (decay length) requirement
is estimated using $|\varepsilon_{\rm data}-\varepsilon_{\rm MC}|/\varepsilon_{\rm data}$,
where $\varepsilon_{\rm data}$ is the efficiency of applying the $\bar\Lambda$ ($\bar\Xi^+$)
mass window (decay length) requirement 
by extracting $\bar\Lambda$ ($\bar\Xi^+$) signal in the $\bar{p}\pi^+$ ($\bar\Lambda\pi^+$) invariant mass spectrum of the data,
and $\varepsilon_{\rm MC}$ is the corresponding efficiency from the MC simulation.
The difference between the data and the MC simulation is
considered as the systematic uncertainty and is found to be 0.2\% (0.1\%)
due to the $\bar\Lambda$ mass window (decay length) requirement, and 1.4\% (1.0\%) for the $\bar\Xi^+$ mass window (decay length) requirement.

\indent 4. Kinematic fit for the radiative $\Xi(1530)^-$ decay mode:
Correcting the tracking helix parameters~\cite{helix} reduces the difference between MC simulation and data.
The uncertainty of 2.4\% due to the kinematic fit is estimated by the observed differences between an analysis that
accounts for such correction and an analysis that does not.
The correction factors obtained by control sample $J/\psi\to p\bar{p}\pi^+\pi^-$ and gives 2.4\% as the estimated systematic uncertainty.

\indent 5. Angular distribution:
The systematic uncertainty of angular distribution is estimated to take the larger difference of the detection efficiency by
varying the measured $\alpha$ values by $\pm1\sigma$ in the MC simulation.
And it is determined to be 0.5\% and 3.6\% for the inclusive and radiative $\Xi(1530)^-$ decay modes, respectively.


\indent 6. Fit procedure: For the inclusive $\Xi(1530)^-$ decay mode,
uncertainties due to the fitting range of $M^{\rm recoil}_{\bar{\Lambda}\pi^+}$ are estimated
by changing the fitting range from 1.47-1.62 GeV/c$^2$ to 1.475-1.615 GeV/c$^2$ and 1.465-1.625 GeV/c$^2$, respectively.
The largest difference with respect to the nominal value is 0.7\% and this is taken as the uncertainty associated with the fitting range.
The uncertainty due to the background shape is estimated by changing the second-order polynomial function to a first-order polynomial. The relative difference on the signal yield of 1.0\% is taken as the uncertainty due to the background shape.
In the fit of $M^{\rm recoil}_{\bar\Lambda\pi^+}$, the signal shape is parametrized
by the simulated MC shape convolved with a Gaussian function with the mean of zero.
To estimate the uncertainty caused by a possible shift of the signal peak,
an alternative model with the free mean of the Gaussian is used
to estimate the uncertainty due to the signal shape.
The difference between the two fits of 0.02\% is negligible.
Assuming that the sources above are independent and adding them in quadrature,
the total systematic uncertainty associated with the fit procedure is obtained to be 1.2\%.
As for the radiative $\Xi(1530)^-$ decay mode, the uncertainty
associated with the fit procedure is negligible since the nominal upper limit on $\mathcal{B}(\Xi(1530)^- \to \gamma \Xi^-)$ is the most conservative one among multiple fit scenarios.

\begin{table}[htbp]
\caption{Systematic uncertainties on the branching fraction measurements. Here, $\Xi^{*-}$ denotes the $\Xi(1530)^-$ resonance.}
\begin{center}
\renewcommand\arraystretch{1.2} 
\vskip -0.2cm
\begin{tabular*}{\columnwidth}{l|c|c}
\hline
\hline
{Source}   & $J/\psi\to\Xi^{*-}\bar\Xi^+(\%)$    & $\Xi^{*-}\to\gamma\Xi^-$(\%) \\
\hline
Photon                               & -         & 1.0 \\
$\bar\Xi^+$ efficiency correction                    & 0.7       & 0.7 \\
$\bar\Lambda/\Lambda$ mass window    & 0.2       & 0.2 \\
$\bar\Xi^+/\Xi^-$ mass window        & 1.4       & 1.4 \\
$\bar\Lambda/\Lambda$ decay length   & 0.1       & 0.1 \\
$\bar\Xi^+/\Xi^-$ decay length       & 1.0       & 1.0 \\
Kinematic fit                        & -         & 2.4 \\
Angular distribution                 & 0.5       & 3.6 \\
Fit procedure                        & 1.2       & - \\
Intermediate decays                  & 0.8       & 0.8 \\
$N_{J/\psi}$                         & 0.5       & -   \\
\hline
{\bf In total }                      & 2.5        & 4.9 \\
\hline
\hline
\end{tabular*}
\end{center}
\label{error3}
\end{table}
\vskip -0.1cm


\indent 8. Intermediate decays:
The uncertainties due to the branching fractions of intermediate decays $\Xi^-\to\Lambda\pi^-$ and $\Lambda\to p\pi^-$ are 0.04\% and 0.8\%~\cite{pdg}, respectively. Therefore, this uncertainty associated with the branching fractions of intermediate decays is taken to be 0.8\%.


\indent 9. Number of $J/\psi$ events:
The total number of $J/\psi$ events is obtained by studying the inclusive hadronic $J/\psi$ decays which has a systematic
uncertainty of 0.5\%~\cite{totJpsi}.



Table~\ref{error3} lists all systematic uncertainties on branching fraction measurements
for the $J/\psi\to\Xi(1530)^-\bar\Xi^+$ decay in the ST mode and the radiative $\Xi(1530)^-$ decay mode, respectively.
The total systematic uncertainty is individually calculated as the quadratic sum of all individual terms for each mode.

\subsection{Angular distribution}
\vspace{-0.4cm}

The systematic uncertainties in the measurement of the $\alpha$ value arise from $M^{\rm recoil}_{\bar\Lambda\pi^+}$ fitting range,  background shape, $\rm cos\theta$ fitting range, $\rm cos\theta$ binning, and efficiency correction.
It should be noted that the absolute value of the difference between the re-measured $\alpha$ values in the alternative cases mentioned above and the nominal value is taken as the uncertainty given in this analysis.

\begin{table}[htbp]
\caption{Absolute systematic uncertainties on the $\alpha$ value.}
\begin{center}
\renewcommand\arraystretch{1.2} 
\begin{tabular*}{0.9\columnwidth}{@{\extracolsep{\fill}}rl}
\hline
\hline
Source & $\alpha_{\Xi(1530)^-}$ \\
\hline
$M^{\rm recoil}_{\bar\Lambda\pi^+}$ fitting range & 0.02 \\
Background shape  &  0.04 \\
cos$\theta$ fitting range & 0.01 \\
cos$\theta$ binning & 0.01 \\
Efficiency correction &0.03\\
\hline
{\bf Total uncertainty}      & 0.06 \\
\hline
\hline
\end{tabular*}
\end{center}
\label{erroralpha}
\end{table}

\indent 1. The $M^{\rm recoil}_{\bar\Lambda\pi^+}$ fitting range:
The uncertainty due to the fitting range of $M^{\rm recoil}_{\bar\Lambda\pi^+}$  is estimated by changing the fitting range from 1.47-1.62 GeV/c$^2$ to 1.475-1.615 GeV/c$^2$ and 1.465-1.625 GeV/c$^2$, respectively. The largest difference for $\alpha_{\Xi(1530)^-}$ of 0.02 is taken as the uncertainty due to the fitting range.

\indent 2. The background shape:
The uncertainty due to the background shape in the angular distribution is estimated by changing the second-order polynomial function applied for fitting $M^{\rm recoil}_{\bar\Lambda\pi^+}$ to a first-order polynomial function. The difference becomes 0.04 for $\alpha_{\Xi(1530)^-}$ and this is taken as the uncertainty due to the background shape.

\indent 3. The $\rm cos\theta$ fitting range:
The uncertainty due to the $\rm cos\theta$ fitting range is estimated by varying the cos$\theta$ fitting range to [-0.9, 0.9]. The difference on angular distribution is 0.01 and
this is taken as the uncertainty due to the $\rm cos\theta$ fitting range.

\indent 4. The cos$\theta$ binning:
The uncertainty due to the binning of cos$\theta$ is estimated by changing the nominal choice of 20 bins to 10 bins.
The difference for $\alpha$ value between the the two cases
 of 0.01 is taken as the systematic uncertainty due to the binning.

\indent 5. Efficiency correction:
The $\alpha$ value is obtained by
fitting the efficiency-corrected cos$\theta$ distribution. To estimate the systematic uncertainty due to
the MC generator to the fitted $\alpha$ value, the ratio of detection efficiencies  between the data and MC simulation
is obtained based on the process $J/\psi\to \Xi(1530)^-\bar\Xi^+$ with
the inclusive decay of $\Xi(1530)^-$.
The cos$\theta$ distribution is  refitted using
corrected one by the above ratio of detection efficiencies. The resulting absolute difference of 0.03 in $\alpha$ is taken as the systematic uncertainty due to
the imperfection of MC simulation.


The absolute systematic uncertainties from the different sources for the $\alpha$ parameter of the angular distribution are given in Table~\ref{erroralpha},
and the total systematic uncertainty is obtained by adding the values in quadrature, assuming that the sources of uncertainty are independent.

\section{Summary and Discussion}
\label{sec:summary}
\vspace{-0.4cm}

\begin{table*}[htbp]
\caption{ Comparison of the results from this measurements to previous work.
}
\begin{center}
\begin{tabular}{c|c|c|c}
\hline
\hline
 & This work & Other measurements & Theoretical prediction \\\hline
${\cal{B}}(J/\psi\to\Xi(1530)^{-}\bar{\Xi}^{+} +c.c.$)~($10^{-4}$) & $3.17\pm0.02\pm0.08$ & $5.9\pm0.9\pm1.2$~\cite{DM2} &-\\ \hline
$\alpha(J/\psi\to\Xi(1530)^{-}\bar{\Xi}^{+} )$ & $-0.21\pm0.04\pm0.06$ & - & - \\ \hline
${\cal{B}}(\Xi(1530)^-\to\gamma\Xi^-)~(\%)$ & $\leq3.7$ @ 90\% C.L. & $\leq$~4 @ 90\% C.L.~\cite{c2} & $\sim 0.03$~\cite{c1,Ramalho:2013uza}\\
\hline
\hline
\end{tabular}
\end{center}
\label{tab:summary}
\end{table*}

The SU(3)-flavor violating decay $J/\psi\to\Xi(1530)^{-}\bar\Xi^{+} $ is
measured using $(1310.6\pm7.0)\times 10^{6}$ $J/\psi$ events collected with the BESIII detector in 2009 and 2012.
The signal is clearly  observed ($ > 10~\sigma$) and the branching fraction is measured to be
${\cal{B}}(J/\psi\to\Xi(1530)^{-}\bar\Xi^{+}+c.c.)=(3.17\pm0.02\pm0.08)\times10^{-4}$.
The result is consistent with the DM2 measurement~\cite{DM2} within
2 standard deviations
(see Table~\ref{tab:summary}),
but with an order of magnitude improved precision.
The $\alpha$ value of the angular distribution for   $J/\psi\to\Xi(1530)^{-}\bar\Xi^{+} $ decay
is measured for the first time and is found to be $\alpha_{\Xi(1530)}=-0.21\pm0.04\pm0.06$.


In addition, we present the first evidence for the $\Xi(1530)^-\to\gamma\Xi^-$ radiative decay with a significance of 3.9$\sigma$.
The upper limit at the 90\% C.L. on the branching fraction of $\Xi(1530)^-\to\gamma\Xi^-  $ is measured to be
3.7\%, which is consistent with the previous measurement~\cite{c2}.
The result is compatible with the theoretical prediction of $3.0\times10^{-4}$~\cite{c1,Ramalho:2013uza}.
Our result provides complementary experimental information for isolating both the octet-decuplet mixing mechanism~\cite{c1} and meson cloud effects~\cite{Ramalho:2013uza} in the baryon structure.


\begin{acknowledgements}
\label{sec:acknowledgement}
\vspace{-0.4cm}
The BESIII Collaboration thanks the staff of BEPCII and the IHEP computing center for their strong support. This work is supported in part by National Key Basic Research Program of China under Contract No. 2015CB856700; National Natural Science Foundation of China (NSFC) under Contracts Nos. 11565006, 11605042, 11625523, 11635010, 11735014, 11835012, 11935018; the Chinese Academy of Sciences (CAS) Large-Scale Scientific Facility Program; Joint Large-Scale Scientific Facility Funds of the NSFC and CAS under Contracts Nos. U1232107, U1532257, U1532258, U1732263, U1832207; CAS Key Research Program of Frontier Sciences under Contracts Nos. QYZDJ-SSW-SLH003, QYZDJ-SSW-SLH040; 100 Talents Program of CAS; INPAC and Shanghai Key Laboratory for Particle Physics and Cosmology; German Research Foundation DFG under Contract No. Collaborative Research Center CRC 1044; Istituto Nazionale di Fisica Nucleare, Italy; Koninklijke Nederlandse Akademie van Wetenschappen (KNAW) under Contract No. 530-4CDP03; Ministry of Development of Turkey under Contract No. DPT2006K-120470; National Science and Technology fund; The Knut and Alice Wallenberg Foundation (Sweden) under Contract No. 2016.0157; The Swedish Research Council; U. S. Department of Energy under Contracts Nos. DE-FG02-05ER41374, DE-SC-0010118, DE-SC-0012069; University of Groningen (RuG) and the Helmholtzzentrum fuer Schwerionenforschung GmbH (GSI), Darmstadt; China Postdoctoral Science Foundation under Contract No. 2017M622347,
Postdoctoral research start-up fees of Henan Province under Contract No. 2017SBH005,
Ph.D research start-up fees of Henan Normal University under Contract No. qd16164,
Program for Innovative Research Team in University of Henan Province ( Grant No.
19IRTSTHN018).

\end{acknowledgements}


\begin{thebibliography}{99}

  \bibitem{su3}
  \href{https://arxiv.org/abs/0809.1869}
  {D. M. Asner {\it et al.}, Int. J. Mod. Phys. A {\bf24}, Supp. 1, 247 (2009), https://doi.org/10.1142/S0217751X09046527.}

  \bibitem{jpsidecay1989}
  \href{https://doi.org/10.1016/0370-1573(89)90074-4}
  {L. K\"{o}pke and N. Wermes, Phys. Rept. {\bf 174}, 67 (1989).}

  \bibitem{Ramalho:2013uza}
  \href{https://doi.org/10.1103/PhysRevD.87.093011}
       {G.~Ramalho and K.~Tsushima, 
       Phys.\ Rev.\ D {\bf 87}, 093011 (2013).}
 \bibitem{R00}
       {Inclusion of charge-conjugate decay mode is implicit unless otherwise stated.}

  \bibitem{Jpsidecay1976}
  \href{10.1103/PhysRevD.14.852}
  {H. Kowalski and T. F. Walsh, Phys. Rev. D {\bf 14}, 852 (1976).}

  \bibitem{R11}{Unless otherwise indicated in the paper, a single number on the uncertainty refers to the sum in quadrature of the statistical item and the systematic one; if there are two uncertainties, the first one is statistical, and the second systematic.}

  \bibitem{DM2}
  \href{https://doi.org/10.1016/0550-3213(87)90664-X}
       {P.~Henrard {\it et al.} (DM2 Collaboration), 
       Nucl.\ Phys.\ B {\bf 292}, 670 (1987). }

 \bibitem{wxf}
       {M.~Ablikim {\it et al.} (BESIII Collaboration), 
       Phys. Rev. D {\bf 100}, 051101(R) (2019).}

  \bibitem{pdg}
  \href{https://doi.org/10.1088/1674-1137/40/10/100001}
  {M. Tanabashi {\em et al.} (Particle Data Group), Phys. Rev. D {\bf 98}, 030001 (2018)}.

  \bibitem{Zhang:2004xt}
  \href{http://hepnp.ihep.ac.cn/article/id/dfda0110-74a4-4b0b-a5c8-a75b86ca9b36}
       {A.~L.~Zhang, Y.~R.~Liu, P.~Z.~Huang, W.~Z.~Deng, X.~L.~Chen and S.~L.~Zhu, 
       Chin. Phys. C {\bf 29}, 250 (2005), http://cpc.ihep.ac.cn/article/id/dfda0110-74a4-4b0b-a5c8-a75b86ca9b36.
       }

  \bibitem{data0}
  \href{https://doi.org/10.1088/1674-1137/36/10/001}
       {M.~Ablikim {\it et al.} (BESIII Collaboration),
       Chin.\ Phys.\ C {\bf 36}, 915 (2012).}
  \bibitem{totJpsi}
  \href{https://doi.org/10.1088/1674-1137/41/1/013001}
       {M.~Ablikim {\it et al.} (BESIII Collaboration), 
       Chin.\ Phys.\ C {\bf 41}, 013001 (2017).}

   \bibitem{Brodsky:1981kj}
   \href{https://doi.org/10.1103/PhysRevD.24.2848}
  {S.~J.~Brodsky and G.~P.~Lepage,
  Phys.\ Rev.\ D {\bf 24}, 2848 (1981).}

   \bibitem{alpha1}
  \href{https://doi.org/10.1103/PhysRevD.25.1345}
       {M.~Claudson, S.~L.~Glashow, and M.~B.~Wise,
       Phys.\ Rev.\ D {\bf 25}, 1345 (1982).}
  \bibitem{alpha2}
  \href{https://doi.org/10.1142/S0217751X87000107}
       {C.~Carimalo, 
       Int.\ J.\ Mod.\ Phys.\ A {\bf 2}, 249 (1987).}
  \bibitem{alpha3}
  \href{https://doi.org/10.1103/PhysRevD.51.3487}
       {F.~Murgia and M.~Melis, 
       Phys.\ Rev.\ D {\bf 51}, 3487 (1995)}.
  \bibitem{alpha4}
  \href{https://doi.org/10.1016/j.physletb.2006.05.093}
  {H.~Chen and R.~G.~Ping, 
  Phys.\ Lett.\ B {\bf 644}, 54 (2007).}

  \bibitem{Ablikim:2005cda}
  \href{https://doi.org/10.1016/j.physletb.2005.10.079}
  {M.~Ablikim {\it et al.} (BES Collaboration),
  Phys.\ Lett.\ B {\bf 632}, 181 (2006).}

  \bibitem{xiongfei1617}
  \href{https://doi.org/10.1103/PhysRevD.93.072003}
  {M.~Ablikim {\it et al.} (BESIII Collaboration),
  Phys.\ Rev.\ D {\bf 93}, 072003 (2016);}
  \href{https://doi.org/10.1016/j.physletb.2017.04.048}
  {M.~Ablikim {\it et al.} (BESIII Collaboration),
  Phys.\ Lett.\ B {\bf 770}, 217 (2017).}


  \bibitem{c1} E. Kaxiras and E. J. Moniz, \ Phys. \ Rev. D {\bf 32}, 695 (1985).
  \bibitem{Myhrer:2006cu}
  \href{https://doi.org/10.1103/PhysRevC.74.065202}
       {F.~Myhrer, 
       Phys.\ Rev.\ C {\bf 74}, 065202 (2006).}

  \bibitem{Li:2016tlt}
  \href{https://doi.org/10.1007/s11467-017-0691-9}
       {H.~B.~Li, 
        Front.\ Phys.\ {\bf 12}, 121301 (2017);}
    \href{https://doi.org/10.1007/s11467-019-0910-7}
     {Erratum: Front.\ Phys.\ {\bf 14}, 64001 (2019).}

  \bibitem{c2} G. R. Kalbfleisch, R. C. Strand, and J. W. Chapman, \ Phys. \ Rev. D {\bf 11}, 987 (1975).


  \bibitem{nima614.345}
  \href{https://doi.org/10.1016/j.nima.2009.12.050}
       {M. Ablikim {\em et al.} (BESIII Collaboration), 
       Nucl. Instrum. Meth. A {\bf 614}, 345 (2010).}
  \bibitem{cpc30.371}
  \href{http://cpc-hepnp.ihep.ac.cn:8080/Jwk_cpc/EN/Y2006/V30/I05/371}
       {Z.~Y.~Deng, G.~F.~Cao, and C.~D.~Fu  {\em et al.}, 
       Chin. Phys. C {\bf 30}, 371 (2006), http://cpc.ihep.ac.cn/article/id/283d17c0-e8fa-4ad7-bfe3-92095466def1}.
  \bibitem{numa506.250}
  \href{https://doi.org/10.1016/S0168-9002(03)01368-8}
       {S. Agostinelli {\em et al.} (GEANT4 Collaboration), 
       Nucl. Instrum. Meth. A {\bf 506}, 250 (2003)}.
  \bibitem{tns53.270}
  \href{https://doi.org/10.1109/TNS.2006.869826}
       {J. Allison {\em et al.}, 
       IEEE Trans. Nucl. Sci. {\bf 53}, 270 (2006)}.
  \bibitem{J2BB3}
  \href{http://bes.ihep.ac.cn/bes3/phy_book/book/phy/generators_ping.pdf.pdf}
       {R.~G.~Ping and C.~Y.~Pang, ``Monte Carlo Generators for Tau-Charm-Physics at BESIII",
       \url{http://bes.ihep.ac.cn/bes3/phy_book/book/phy/generators_ping.pdf.pdf}.}


  \bibitem{kkmc}
  \href{https://doi.org/10.1016/S0010-4655(00)00048-5}
       {S. Jadach, B.~F.~L.~Ward, and Z. Was, 
       Comput. Phys. Commun. {\bf 130}, 260 (2000)};
  \href{https://doi.org/10.1103/PhysRevD.63.113009}
       {S.~Jadach, B.~F.~L.~Ward, and Z.~Was,
       Phys. Rev. D {\bf 63}, 113009 (2001)}.


  \bibitem{evtgen}
  \href{https://doi.org/10.1016/S0168-9002(01)00089-4}
       {D. Lange, 
       Nucl. Instrum. Meth. A  {\bf 462}, 152 (2001)};
  \href{https://doi.org/10.1088/1674-1137/32/8/001}
       {R.~G.~Ping, 
       Chin. Phys. C {\bf 32}, 599 (2008)}.

  \bibitem{lundcharm}
  \href{https://doi.org/10.1103/PhysRevD.62.034003}
       {J.~C.~Chen, G. S. Huang, X. R. Qi, D. H. Zhang, and Y. S. Zhu,
       Phys. Rev. D {\bf 62}, 034003 (2000)}.

  \bibitem{sec-vtx}
  \href{https://doi.org/10.1088/1674-1137/33/6/005}
       {M.~Xu {\it et al.},
       Chin.\ Phys.\ C {\bf 33}, 428 (2009).}


\bibitem{Wilk}
       {S.~I. Bityukov and N. ~V. Krasnikov,
       Nucl. Instrum. Meth. A{\bf 452}, 518 (2000)}.

  \bibitem{gamrecon}
  \href{https://doi.org/10.1103/PhysRevD.83.112005}
  {M.~Ablikim {\it et al.} (BESIII Collaboration),
  Phys.\ Rev.\ D {\bf 83}, 112005 (2011).}

  \bibitem{helix}
  \href{https://doi.org/10.1103/PhysRevD.87.012002}
  {
  M.~Ablikim {\it et al.} (BESIII Collaboration),
  Phys.\ Rev.\ D {\bf 87}, 012002 (2013).
  }


\end{thebibliography}
\end{document}